\documentclass[letterpaper]{article} 
\usepackage[preprint]{aaai2027}  
\usepackage[hyphens]{url}  
\usepackage{graphicx} 
\usepackage{natbib}  
\usepackage{caption} 
\usepackage{algorithm}
\usepackage{algorithmic}
\usepackage{booktabs}
\usepackage{amsmath}
\usepackage{amssymb}

\usepackage{xcolor}
\definecolor{darkblue}{rgb}{0, 0, 0.5}
\usepackage[breaklinks=true]{hyperref}
\hypersetup{colorlinks=true, citecolor=darkblue, linkcolor=darkblue, urlcolor=darkblue}

\title{Restoring Collaborative Signals in Semantic-ID Generative Recommendation via Personalized Natural Language}

\author{
    Changjiang Han\textsuperscript{1,2},
    Qingyang Li\textsuperscript{2},
    Yaqiang Zang\textsuperscript{2},
    Jikun Kang\textsuperscript{3},\\
    Pinghua Gong\textsuperscript{2},
    Xue Liu\textsuperscript{1,4,5},
    Bowei He\textsuperscript{1,4}\thanks{Corresponding author: bowei.he@mbzuai.ac.ae.}
}
\affiliations{
    \textsuperscript{1}Mohamed bin Zayed University of Artificial Intelligence (MBZUAI)\\
    \textsuperscript{2}JD.com\quad
    \textsuperscript{3}Salesforce\quad
    \textsuperscript{4}McGill University\quad
    \textsuperscript{5}Mila -- Quebec AI Institute
}

\begin{document}

\maketitle

\begin{abstract}
Making LLM-based generative recommendation models stronger and more personalized through natural language and explicit reasoning is a widely anticipated yet still unsolved goal. Such models cast recommendation as autoregressively generating an item's semantic-ID (SID), a short tuple of discrete codes, so that recommending well reduces to emitting the right SID. In this setting the model verbalizes its knowledge poorly, and text and SID tokens live in misaligned embedding spaces. Deep reasoning therefore rarely turns into a correct SID, and enabling explicit ``thinking'' often gives no gain or even hurts. The deeper cause is that a compact SID cannot hold content and collaborative signal at once: the two compete, and collaboration loses. Because a mis-predicted SID is a wrong recommendation, this caps accuracy directly. Costly multi-round training barely helps, and few methods try to supply the missing signal at inference time. What is missing is a reliable channel that carries collaborative signal into SID generation. We therefore propose a framework, guided by personalized natural language, that adds hierarchical collaborative cues as the model generates, without altering the backbone or retraining the SIDs. Rather than mapping language onto SIDs directly, it uses language to attach analyzable links between collaborative patterns and their audiences, restoring the collaborative signal that SIDs miss. The result is consistent gains in recommendation accuracy, grounding generation in collaborative structure at inference time rather than relying on explicit reasoning or retraining.

\end{abstract}


\section{Introduction}

Generative recommendation has emerged as a compelling alternative to the traditional retrieve-and-rank pipeline. Instead of scoring a fixed catalogue of atomic item identifiers, models such as TIGER~\cite{tiger} and OneRec~\cite{onerec} represent each item as a short sequence of discrete codes, a \emph{semantic ID} (SID), and cast recommendation as autoregressive generation over these codes. Because SIDs are derived from item content, they promise better generalization to cold and long-tail items, along with a shared vocabulary between recommendation and natural language. This shared vocabulary has fuelled an optimistic expectation: that natural language (NL) and explicit reasoning can make an SID generator stronger, more controllable, and more personalized, much as chain-of-thought has advanced language tasks~\cite{cot}.

\begin{figure}[t]
\centering
\includegraphics[width=\columnwidth]{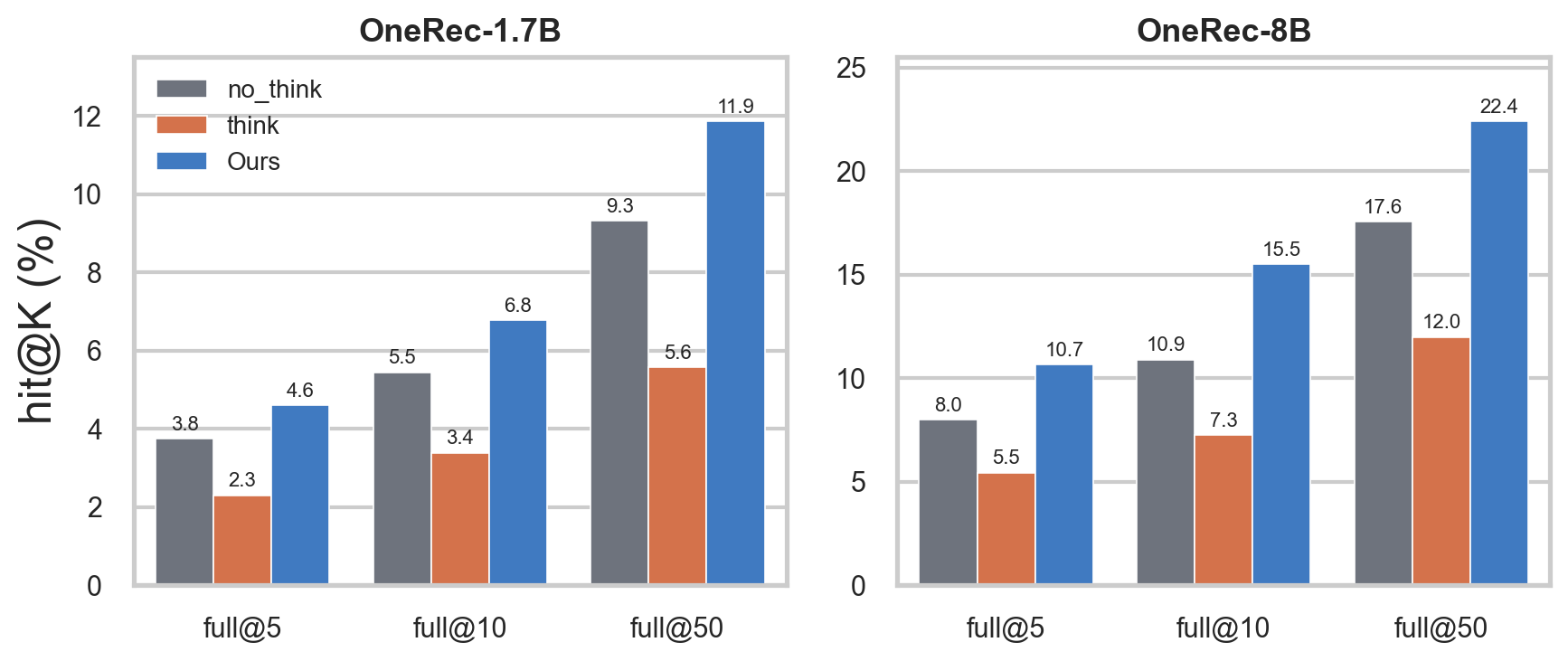}
\caption{Full-SID hit@$K$ on OneRec-1.7B/8B for direct decoding (\textsc{no\_think}), native reasoning (\textsc{think}), and our reranking (\textsc{Ours}); all share the frozen backbone and decoding budget. A hit requires generating the target's exact three-level ID $(s_a,s_b,s_c)$ within top $K$. Enabling reasoning lowers accuracy while \textsc{Ours} raises it.}
\label{fig:thinkablation}
\end{figure}

\begin{figure}[t]
\centering
\includegraphics[width=\columnwidth]{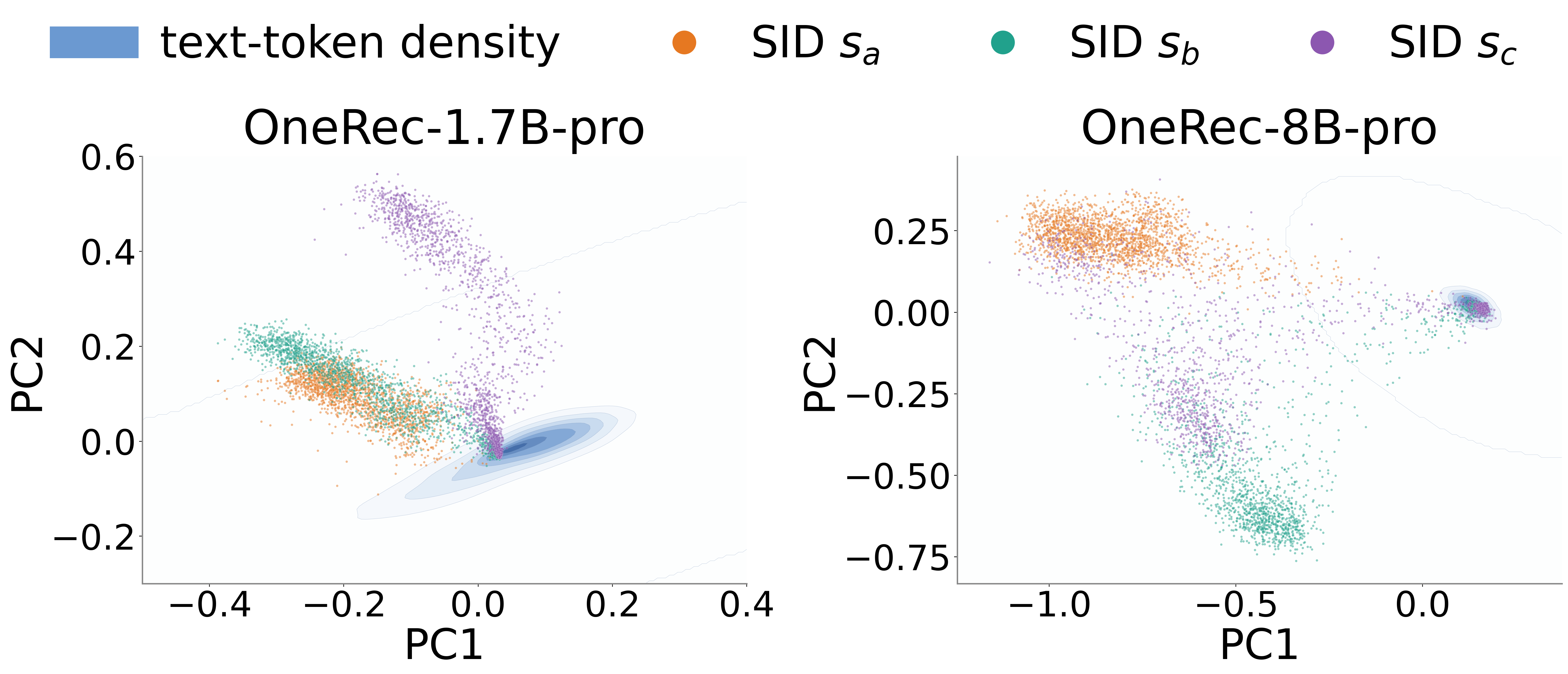}
\caption{PCA of text-token vs.\ three-level SID-token embeddings in the frozen OneRec-pro backbone (1.7B, 8B). About $75$--$79\%$ of SID tokens fall outside the $90\%$-density region of text tokens. Top two components shown; separation is quantified in full embedding dimension.}
\label{fig:embedsep}
\end{figure}

That expectation remains largely unmet. In practice, information produced by deep reasoning rarely converts into a correct SID, and simply enabling an explicit ``thinking'' mode often yields no gain or even hurts~\cite{onereason,implicitreason} (Figure~\ref{fig:thinkablation}). This mirrors the broader finding that reasoning traces need not be faithful to the computation that determines the answer~\cite{cotunfaithful}. A basic reason is representational: the model verbalizes its knowledge poorly, and text and SID tokens live in misaligned embedding spaces (Figure~\ref{fig:embedsep}). The real obstacle, we argue, is the lack of a reliable channel through which natural-language preferences can reach the SID generation decision.

Why does a reliable channel not already exist? The answer comes down to what an SID can carry. Mainstream SIDs are built by residual quantization of content embeddings~\cite{tiger,onerec}, with collaborative signals folded in, if at all, only as an optional auxiliary~\cite{lcrec,letter}. Encoding semantic content and encoding collaborative structure then compete for the same small code, and recent work shows the two impose conflicting optimization objectives that a single SID cannot satisfy at once~\cite{discrec}. Collaboration loses this competition: co-occurrence structure that a classical collaborative filter captures easily~\cite{mf,ncf} is left under-expressed in the code. Reasoning in NL cannot recover it, because the missing quantity was never a linguistic fact but a statistical regularity over behaviour.

Two escape routes have been explored, and both are unsatisfying. The first is to \emph{train harder}: multi-round, reasoning-augmented post-training that tries to internalize collaborative competence~\cite{onerecthink,onereason}. This is expensive and, on open backbones, still struggles to make the generator use collaborative information. The second is to inject signals at inference time; but existing training-free approaches largely re-rank with LLM-derived semantic similarity~\cite{querec}, which does not target the collaborative gap, and such attempts stay scarce.

We take the second route but target the collaborative gap directly. Instead of describing the preference in language and hoping the model uses it, we turn it into a signal that acts on the SID decision. Since the missing co-consumption structure is not something natural language can address on its own, our idea is to make it addressable: we recover it offline as a collaborative factor space, learn a bridge $P(f\mid t)$ that maps a user's natural-language audience tags (summarized from history by an LLM) into that space, and, over a \emph{frozen} backbone, let the resulting query rerank the model's own candidates level by level under a fixed beam. This gives natural-language preferences a path into SID generation that prior text-interface routes lack.

Our contributions are:
\begin{itemize}
    \item \textbf{Diagnosis.} We trace the failure of natural language and reasoning on SID recommendation to a representational cause: the content--collaborative exclusivity of compact SIDs~\cite{discrec}, under which visually similar but audience-opposed items collide at the coarse SID level. We also show that enabling the backbone's own reasoning consistently lowers accuracy.
    \item \textbf{Method.} We give a training-free channel that supplies the missing collaborative signal in four steps. (i) We recover the collaborative structure the SID drops by factorizing a second-order item--item co-view matrix into per-item collaborative factors. (ii) An LLM reads a user's history and summarizes it into natural-language audience tags. (iii) A learned semantic bridge maps these tags into the same collaborative factor space, turning the language description into a query the generator can act on. (iv) At inference, over a frozen backbone, this query is injected level by level into SID generation: it expands the coarse-code candidates at $s_a$, reranks the audience-level candidates at $s_b$, and constrains $s_c$ to valid inventory. The bridge is faithful (the language-reconstructed query matches the behavioural one) and leakage-free.
    \item \textbf{Results.} On two open backbones the channel yields consistent hit-rate gains (OneRec-8B hit@10 $11.14\!\to\!15.50$); per-level attribution shows the gain is carried at the mid-level $s_b$ decision, and it depends on the specific collaborative representation rather than an arbitrary low-dimensional one. The language-reconstructed query differentiates users' audiences at the ranking level ($+8.3$ percentile points over a different-user query, $p<10^{-4}$); this differentiation is real but modest in effect size, which we attribute to the current profile's granularity rather than to the channel.
\end{itemize}

\section{Related Work}

\paragraph{Semantic-ID recommendation and collaborative tokenization.}
TIGER~\cite{tiger} established the standard recipe, encoding item content, quantizing it into discrete codes with an RQ-VAE~\cite{rqvae}, and generating the codes autoregressively, building on earlier work casting recommendation as language-style generation~\cite{p5}; semantic IDs improve generalization to cold and long-tail items~\cite{semanticid}, and OneRec~\cite{onerec} scales the paradigm to industry with a residual $K$-means quantizer. Because SIDs are content-derived, several works inject collaborative signal into the tokenizer: LC-Rec~\cite{lcrec} aligns an LLM's semantic space with collaborative semantics and LETTER~\cite{letter} adds contrastive collaborative regularization, while DiscRec~\cite{discrec} argues the two signals impose conflicting objectives and \emph{disentangles} them. We share DiscRec's diagnosis but draw the opposite conclusion: rather than retrain the tokenizer, we leave the SID untouched and supply the missing signal externally at inference. That signal is co-consumption structure, which classical collaborative filtering captures easily~\cite{mf,bpr,ncf} but a content code under-expresses (cf.\ evidence that such structure is linearly recoverable from language representations~\cite{langrep}).

\paragraph{LLM- and reasoning-based recommendation.}
A broad line adapts LLMs to recommendation through fine-tuning: TALLRec~\cite{tallrec} and LLaRA~\cite{llara} instruction-tune, CoLLM~\cite{collm} injects collaborative embeddings, and RecFormer~\cite{recformer} learns language representations for sequential recommendation~\cite{llmrecsurvey}. A parallel line adds explicit reasoning, mirroring chain-of-thought prompting~\cite{cot,selfconsistency}: OneReason~\cite{onereason} and OneRec-Think~\cite{onerecthink} add reasoning scaffolding through extra training, yet naive ``thinking'' brings little gain for SID prediction, which PauseRec~\cite{implicitreason} attributes to a poor interface between free-form rationales and SID generation, consistent with rationales being unfaithful to the model's computation~\cite{cotunfaithful}. All of these update model parameters; we take the ineffectiveness of inference-time reasoning as motivation and keep the generator frozen.

\paragraph{Inference-time and training-free injection.}
A few approaches add signal without updating the recommendation model. QueRec~\cite{querec} is training-free: it prompts an LLM to generate personalized queries and fuses LLM semantic scores with collaborative-filter scores by re-ranking. Such methods, however, largely exploit \emph{semantic} similarity and do not target the collaborative deficit inside SIDs, and inference-time injection of specifically collaborative structure into a generative SID recommendation model remains little explored. We fill this gap: our guidance derives from second-order co-view structure via hierarchical Poisson factorization~\cite{hpf}, is attached to items through natural language, and re-ranks the backbone's own in-pool candidates.

\section{Methodology}

\paragraph{Setting.}
Each item $i$ is mapped to a semantic ID $s(i)=(s_a,s_b,s_c)$, a three-level tuple of discrete codes produced by residual quantization~\cite{rqvae} of the item's content embedding~\cite{tiger,onerec}; the levels are hierarchical, so items sharing a prefix are content-similar. A frozen backbone $p_\theta$ models the next item from history $\mathcal{H}_u$ autoregressively, $p_\theta(s\mid\mathcal{H}_u)=p_\theta(s_a\mid\mathcal{H}_u)\,p_\theta(s_b\mid s_a,\mathcal{H}_u)\,p_\theta(s_c\mid s_a,s_b,\mathcal{H}_u)$, and recommends by constrained beam search; we write $\ell_a,\ell_b,\ell_c$ for the code log-probabilities. We keep $\theta$ and the SIDs fixed and act only on the scoring of candidates the backbone already proposes, using no ground-truth item information.

\begin{figure*}[t]
\centering
\includegraphics[width=\textwidth]{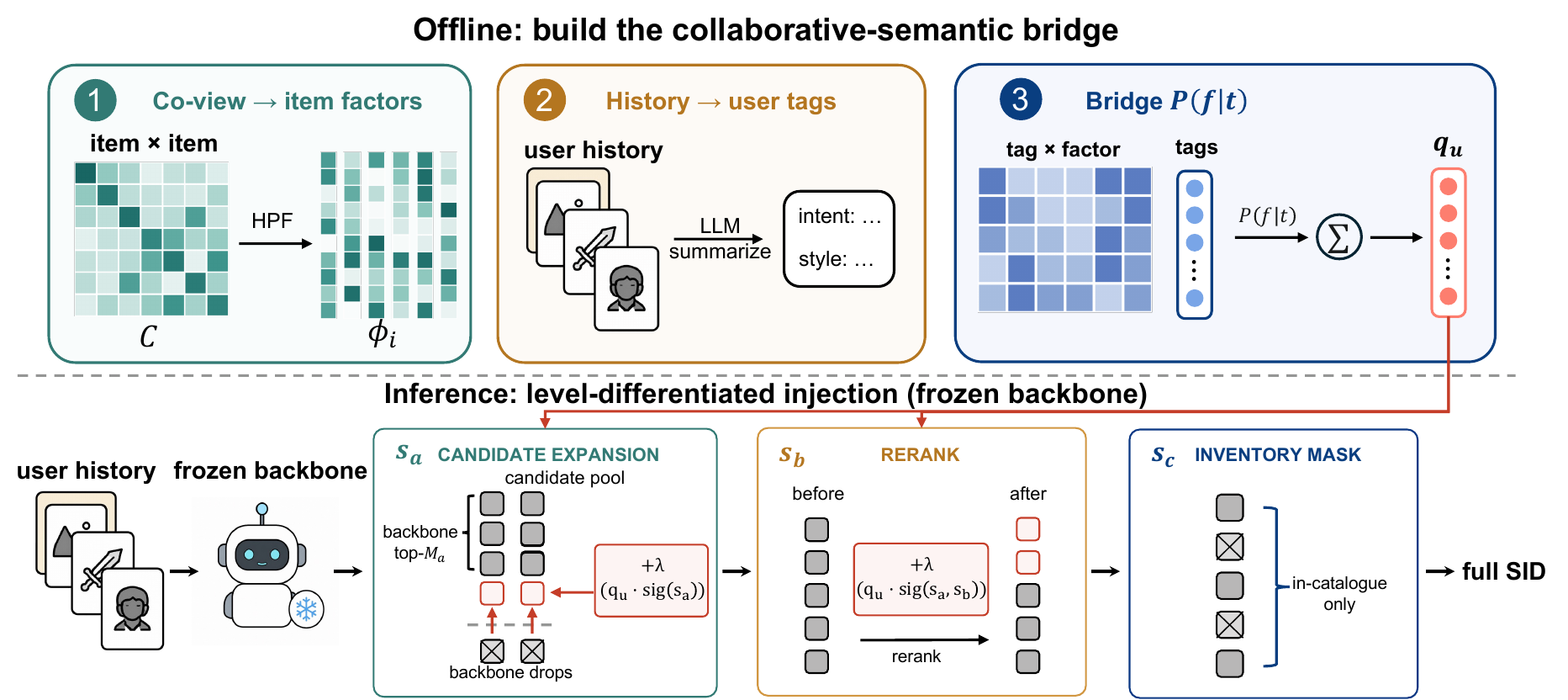}
\caption{Overview. \textbf{Offline (top)}: co-view counts $C$ are factorized with HPF into item factors $\phi_i$; an LLM turns each history into audience/behaviour tags; a bridge $P(f\mid t)$ maps tags into the factor space and sums them into a user vector $q_u$. \textbf{Inference (bottom)}: over a frozen backbone, $q_u$ enters each level differently without widening the beam, expanding $s_a$ candidates, reranking $s_b$, and inventory-masking $s_c$. Coral marks the collaborative residual.}
\label{fig:overview}
\end{figure*}

Our framework has three parts (Figure~\ref{fig:overview}): (i) a behaviour-derived collaborative structure over items, obtained by second-order hierarchical Poisson factorization; (ii) a natural-language bridge that maps a user's LLM-summarized audience tags into this collaborative factor space; and (iii) a level-differentiated, inference-time reranking rule that re-orders the frozen backbone's own candidate codes under a fixed beam budget, adding a collaborative residual at each SID level. Parts (i)--(ii) are precomputed offline; part (iii) runs during decoding and never touches the backbone weights or the SID codebooks.

\subsection{Second-Order Collaborative Semantic Bridge}
\label{sec:bridge}

The central move of our method is to build a \emph{second-order} collaborative structure that bypasses the content signal baked into SIDs. Content-built SIDs place two items close when their \emph{content} embeddings are close; they under-represent the orthogonal fact that two items are \emph{co-consumed} by the same audience. We recover this fact and attach it back to the SID space as a per-prefix collaborative signature.

\paragraph{From first-order to second-order co-occurrence.}
Let $\mathcal{U}$ be the users and $\mathcal{I}$ the items. A first-order (user--item) factorization tends to absorb popularity rather than audience structure, so we build a \emph{second-order} item--item co-view matrix $C \in \mathbb{Z}_{\ge 0}^{|\mathcal{I}|\times|\mathcal{I}|}$,
\begin{equation}
    C_{ij} = \big|\{\, u \in \mathcal{U} : i,j \in \mathcal{H}_u,\; i\neq j \,\}\big|,
\end{equation}
the number of users who consumed both $i$ and $j$. $C$ is sparse and count-valued, which makes a Gamma--Poisson model a natural fit.

\paragraph{Factorization and prefix signatures.}
We factorize $C$ with hierarchical Poisson factorization (HPF)~\cite{hpf}, a Gamma--Poisson model that gives each item $i$ a nonnegative latent signature $\phi_i\in\mathbb{R}_{\ge0}^{d}$ ($d{=}64$) fit by coordinate-ascent variational inference. Since the backbone emits codes rather than item ids, we aggregate item signatures onto SID prefixes: for each occupied prefix $(s_a,s_b)$ with item set $\mathcal{I}_{ab}=\{\,i : s(i){=}(s_a,s_b,\cdot)\,\}$,
\begin{equation}
    \mathrm{sig}(s_a,s_b) \;=\; \frac{1}{|\mathcal{I}_{ab}|}\sum_{i\in\mathcal{I}_{ab}} \phi_i,
\end{equation}
and analogously a coarse signature $\mathrm{sig}(s_a)$ over all items under code $s_a$. This makes the collaborative structure addressable by exactly the codes the generator produces: a content-agnostic, behaviour-derived vector indexed by the content-derived SID prefix. We stop at the $(s_a,s_b)$ level and build no signature at $s_c$: deeper prefixes hold far fewer items, so their co-view statistics are too sparse to estimate reliably, and the items that remain under a fixed $(s_a,s_b)$ are already so similar that their collaborative factors barely differ, leaving no audience distinction for a signature to capture. The mid-level prefix is thus the finest granularity at which collaborative structure is both well-estimated and discriminative. Algorithm~\ref{alg:bridge} summarizes this offline construction.

\begin{algorithm}[t]
\caption{Offline second-order collaborative bridge}
\label{alg:bridge}
\textbf{Input}: interaction histories $\{\mathcal{H}_u\}$; SID map $s(\cdot)$; rank $d$\\
\textbf{Output}: prefix signatures $\{\mathrm{sig}(s_a,s_b)\}$
\begin{algorithmic}[1]
\STATE $C \gets \mathbf{0}$
\FORALL{$u\in\mathcal{U}$}
    \FORALL{unordered pairs $\{i,j\}\subseteq \mathcal{H}_u$}
        \STATE $C_{ij} \mathrel{+}= 1$;\; $C_{ji}\mathrel{+}=1$
    \ENDFOR
\ENDFOR
\STATE $\{\phi_i\} \gets \textsc{HPF-CAVI}(C, d)$ \COMMENT{Gamma--Poisson fit}
\FORALL{occupied prefixes $(s_a,s_b)$}
    \STATE $\mathrm{sig}(s_a,s_b) \gets \mathrm{mean}\{\phi_i : s(i)=(s_a,s_b,\cdot)\}$
\ENDFOR
\STATE \textbf{return} $\{\mathrm{sig}(s_a,s_b)\}$
\end{algorithmic}
\end{algorithm}

\subsection{Natural-Language Bridge to the Factor Space}
\label{sec:nl}

The query $q_u$ can be read directly from behaviour, as the mean factor of a user's history items, but this couples it to raw history and offers no language handle. We instead learn a bridge that reconstructs $q_u$ from a user's natural-language tags, so the same factor query can be addressed, inspected, and edited in language.

\paragraph{Tags.}
An LLM summarizes each user's history into a small set of canonical tags of two types, \emph{audience intent} (e.g.\ hardcore-competitive, female-oriented) and \emph{behaviour style} (e.g.\ single-IP-focused, broad-interest); raw tags are mapped to a fixed ontology by an exact/synonym table.

\paragraph{Learning the bridge $P(f\mid t)$.}
On the training users we learn, for each tag $t$, a distribution over the collaborative factors. Writing $q^F_u$ for a training user's behavioural factor query (the row-normalized mean of their history-item factors $\phi_i$) and $T_u$ for their tags, we accumulate each user's factors, split evenly across their tags, and smooth toward the global factor prior $\pi$:
\begin{equation}
    P(f\mid t) = \frac{\sum_{u:\,t\in T_u} \tfrac{1}{|T_u|}\,q^F_u \;+\; \mu\,\pi}{\big\lVert\,\cdot\,\big\rVert_1},\qquad
    \pi = \tfrac{1}{|\mathcal{U}_{\mathrm{tr}}|}\sum_u q^F_u,
\end{equation}
with a single smoothing constant $\mu$. Intuitively, a tag is placed at the average collaborative location of the users who carry it.

\paragraph{Serving.}
At inference we never read the target user's history items. Given their tags $T_u$, the natural-language query is the normalized sum of the corresponding tag distributions,
\begin{equation}
    q_u \;=\; \operatorname{normalize}\!\Big(\textstyle\sum_{t\in T_u} P(f\mid t)\Big),
\end{equation}
which lies in the same factor space as the behavioural query and is used identically by \S\ref{sec:steer}. We compare the two queries, natural-language ($q_u$ above) versus the raw-history factor mean, as a control in the experiments.

\subsection{Level-Differentiated Inference-Time Reranking}
\label{sec:steer}

We do \emph{not} widen the beam: at each level we only re-order the candidates the backbone already proposes, adding a collaborative residual controlled by one scalar $\lambda$ (we use $\lambda{=}0.1$, fixed). Let $q_u\in\mathbb{R}_{\ge0}^{d}$ be the user's factor query (\S\ref{sec:nl}) and $r(\cdot)=\lambda\log(\max(q_u^\top\mathrm{sig}(\cdot),0)+\epsilon)$ the collaborative affinity of a code. At the coarse level we add the (softmax-normalized) affinity to the $s_a$ log-probabilities and keep the top $M_a$:
\begin{align}
    S_a(s_a) &= \ell_a(s_a) + \lambda\log\!\big(\mathrm{softmax}_{s_a}\!\max(q_u^\top\mathrm{sig}(s_a),0)+\epsilon\big), \notag\\
    \mathcal{A} &= \operatorname*{top\text{-}}{M_a}\, S_a . \label{eq:lvl-a}
\end{align}
For each retained $s_a\in\mathcal{A}$ we run the backbone's $s_b$ forward pass and rerank its $s_b$ candidates with $r(s_a,s_b)$, carrying the coarse score forward:
\begin{align}
    S_{ab}(s_a,s_b) &= S_a(s_a) + \ell_b(s_b\mid s_a) + r(s_a,s_b), \\
    \mathcal{B} &= \bigcup_{s_a\in\mathcal{A}} \operatorname*{top\text{-}}{M_b}\, S_{ab}(s_a,\cdot) . \label{eq:lvl-b}
\end{align}
Finally $s_c$ is decoded with no collaborative term, under an inventory constraint restricting $s_c$ to in-catalogue completions $\mathrm{Inv}(s_a,s_b)$; the cumulative score ranks the full SIDs:
\begin{equation}
    S_{abc}(s_a,s_b,s_c) = S_{ab}(s_a,s_b) + \ell_c(s_c\mid s_a,s_b). \label{eq:final-score}
\end{equation}
The residual thus enters only at $s_a$ and $s_b$; scores propagate into $S_{abc}$ without renormalization, and the pool stays at most $M_a\times M_b$ regardless of $\lambda$.

\paragraph{Why each level differs.}
The three levels play distinct roles. At $s_a$ the residual acts as \emph{candidate expansion}: reordering by $S_a$ pulls coarse codes the beam would have dropped into the top-$M_a$ set, which on its own does not improve accuracy (\S\ref{sec:exp-mech}) but supplies the candidates the next level reranks. At $s_b$ it is \emph{load-bearing residual reranking}: content-built codes can place audience-opposed but look-alike items under the same coarse code, and the second-order co-view structure separates them, promoting the $s_b$ whose audience matches the user within the fixed budget. At $s_c$ we add no collaborative term, since items under a fixed $(s_a,s_b)$ share nearly identical factors, and instead apply the inventory mask. Reranking adds only an $O(d)$ dot product per candidate and no extra forward pass over the frozen backbone: per case the backbone runs one $s_a$ forward, a fresh $s_b$ forward for each of the retained coarse codes (median $22$ on 1.7B, $27$ on 8B), and an $s_c$ forward over the carried prefixes (median $56$--$58$), the same passes its own cascaded decoding performs. Because it reranks in-pool candidates under a fixed budget, the method's ceiling is how often the target's $(s_a,s_b)$ prefix already lies in $\mathcal{B}$. Algorithm~\ref{alg:steer} summarizes the procedure; the offline bridge construction is given in the supplement.

\begin{algorithm}[t]
\caption{Level-differentiated collaborative reranking (frozen backbone, fixed beam)}
\label{alg:steer}
\textbf{Input}: user factor query $q_u$; backbone log-probs $\ell_a,\ell_b,\ell_c$; signatures $\mathrm{sig}(\cdot)$; budgets $M_a,M_b$; weight $\lambda$\\
\textbf{Output}: ranked SIDs
\begin{algorithmic}[1]
\STATE $S_a \gets \ell_a + \lambda\log(\mathrm{softmax}\,\max(q_u^\top\mathrm{sig}(s_a),0)+\epsilon)$
\STATE $\mathcal{A} \gets \operatorname*{top\text{-}}{M_a}\, S_a$ \COMMENT{$s_a$ candidate expansion}
\STATE $\mathcal{B} \gets \varnothing$
\FORALL{$s_a \in \mathcal{A}$}
    \STATE $S_{ab}(s_a,\cdot) \gets S_a(s_a) + \ell_b(\cdot\mid s_a) + r(s_a,\cdot)$
    \STATE $\mathcal{B} \gets \mathcal{B} \cup \operatorname*{top\text{-}}{M_b}\, S_{ab}(s_a,\cdot)$ \COMMENT{$s_b$ residual reranking}
\ENDFOR
\STATE $\mathcal{S} \gets \{(s_a,s_b,s_c): (s_a,s_b)\in\mathcal{B},\, s_c\in\mathrm{Inv}(s_a,s_b)\}$ \COMMENT{inventory}
\STATE rank $\mathcal{S}$ by $S_{abc}=S_{ab}+\ell_c$ \COMMENT{cumulative, Eq.~\ref{eq:final-score}}
\STATE \textbf{return} top-$K$ of $\mathcal{S}$
\end{algorithmic}
\end{algorithm}

\section{Experiments}

We ask whether natural-language preferences can be compiled into the hierarchical decision process of a frozen SID generator, where the resulting gains arise, and whether the language-reconstructed query differentiates users' audiences. Our findings are: factor-space guidance improves frozen SID decoding; the gain is concentrated at the mid-level $s_b$ decision and depends on the collaborative representation; and the natural-language bridge yields a user-conditioned query that both preserves the downstream utility of a history-derived one and carries a real, if currently modest, user-specific audience signal.

\subsection{Experimental Setup}

\paragraph{Data and evaluation.}
We evaluate on RecIF, the instruction-following recommendation benchmark released with OpenOneRec~\cite{openonerec}, whose test split contains $1{,}000$ next-item prediction cases spanning five recommendation subtasks (interactive, video, ad, label-conditioned, and product recommendation); each case gives a user's interaction history and asks the model to generate the target item's semantic ID. To measure genuine generalization rather than memorized training targets, we hold out every case whose target item also appears as a training prediction target, leaving a GT-disjoint evaluation set of $826$ cases. Statistics for the collaborative bridge come only from training interactions, and the test target is never revealed to the query builder. Unless stated we report full-SID hit@$K$ and NDCG@10 as percentages, with paired-bootstrap $95\%$ confidence intervals ($2000$ resamples). Natural-language queries cover most cases; uncovered ones fall back to the backbone.

\paragraph{Backbones.}
We evaluate two frozen scales from the OneRec family, OneRec-1.7B and 8B; this probes robustness to scale, not cross-architecture generality.

\paragraph{Inference protocol.}
All methods share one harness and candidate budget: the backbone's top-$M_a{=}20$ coarse codes, up to $M_b{=}20$ mid codes per retained $s_a$, top-$40$ prefixes carried to $s_c$, and an inventory-constrained $s_c$ decode; the collaborative residual uses a single fixed weight $\lambda{=}0.1$ (not tuned on test). \textsc{Ours} denotes the full method including the $s_c$ inventory constraint (\S\ref{sec:steer}); all reranking variants operate on the same in-pool candidates from a single forward pass per case, so they add no GPU cost over \textsc{Base} beyond an $O(d)$ dot product. Bold marks the best per column.

\paragraph{Baselines.}
All baselines share the frozen backbone and differ only in the inference-time signal used to rerank its candidates: \textsc{Base} (no external guidance); \textsc{Sem} (caption/text semantic-similarity reranking, a QueRec-style route~\cite{querec}); \textsc{Hist} (a collaborative factor query read from the raw interaction history); and \textsc{Ours} (the same query reconstructed from natural-language tags, \S\ref{sec:nl}). We also run a popularity-reranking control (\textsc{Pop}) under the identical protocol; it stays within a point of \textsc{Base} (1.7B hit@10 $5.08$, 8B $10.90$), so the gains below are not a popularity artifact.

\subsection{Factor-Space Guidance Improves Frozen SID Decoding}
\label{sec:exp-main}

Collaborative factor-space guidance consistently improves both OneRec scales (Table~\ref{tab:main}). On the 8B model, \textsc{Ours} raises hit@10 from $11.14$ to $15.50$, whereas direct semantic reranking (\textsc{Sem}) stays close to the frozen backbone. The funnel columns locate the effect: the coarse-code reachability $s_a$@20 is essentially unchanged across arms (all within a point), while the coarse--mid reachability $(s_a,s_b)$@20 rises markedly under \textsc{Hist} and \textsc{Ours} ($18.2\!\to\!22.0$ on 8B), and this is what carries into full-SID accuracy. The improvement thus comes from reranking in the collaborative factor space at the mid level, not from surface content similarity. \textsc{Hist} and \textsc{Ours} achieve closely matched performance throughout, indicating that the natural-language bridge preserves the predictive utility of the history-derived query. We next examine where this improvement arises and whether the injected query differentiates users.

\begin{table*}[t]
\centering
\setlength{\tabcolsep}{6pt}
\begin{tabular}{@{}lccccccc c@{}}
\toprule
& \multicolumn{3}{c}{Funnel reachability} & \multicolumn{5}{c}{Full-SID accuracy} \\
\cmidrule(lr){2-4}\cmidrule(lr){5-9}
Method & $s_a$@20 & $(s_a,s_b)$@10 & $(s_a,s_b)$@20 & hit@5 & hit@10 & hit@20 & hit@50 & NDCG@10 \\
\midrule
\multicolumn{9}{@{}l}{\emph{OneRec-1.7B}} \\
\textsc{Base} & 43.70 & 9.32 & 12.35 & 2.66 & 4.60 & 7.26 & 8.84 & 2.33 \\
\textsc{Sem}  & \textbf{44.19} & 9.20 & 12.83 & 3.03 & 5.08 & 7.75 & 9.93 & 2.53 \\
\textsc{Hist} & 43.95 & \textbf{11.86} & \textbf{15.74} & \textbf{4.60} & 6.66 & 8.72 & 11.62 & 3.49 \\
\textsc{Ours} & 43.46 & 11.74 & \textbf{15.74} & \textbf{4.60} & \textbf{6.78} & \textbf{8.84} & \textbf{11.86} & \textbf{3.60} \\
\midrule
\multicolumn{9}{@{}l}{\emph{OneRec-8B}} \\
\textsc{Base} & 45.28 & 13.92 & 18.16 & 7.38 & 11.14 & 13.92 & 18.40 & 6.30 \\
\textsc{Sem}  & \textbf{45.40} & 14.65 & 18.89 & 7.51 & 11.86 & 14.89 & 19.73 & 6.96 \\
\textsc{Hist} & 45.04 & \textbf{19.01} & 21.91 & \textbf{10.65} & 15.38 & 18.28 & \textbf{22.52} & 9.38 \\
\textsc{Ours} & \textbf{45.40} & 18.89 & \textbf{22.03} & \textbf{10.65} & \textbf{15.50} & \textbf{18.40} & 22.40 & \textbf{9.59} \\
\bottomrule
\end{tabular}
\caption{Main results (\%). All arms share one harness and candidate pool, differing only in the inference-time reranking signal. $s_a$@20 and $(s_a,s_b)$@$\{10,20\}$ are prefix reachability within top $K$; hit@$K$/NDCG@10 are full-SID accuracy. \textsc{Ours} includes the $s_c$ inventory constraint. Best per column \textbf{bold}.}
\label{tab:main}
\end{table*}

\subsection{The Gain Arises at the Mid-Level SID Decision}
\label{sec:exp-mech}

\paragraph{Level-wise attribution.}
Adding the components one at a time localizes the gain to the mid-level $s_b$ decision (Table~\ref{tab:cum}). Expanding the $s_a$ candidates alone moves accuracy within noise, since it changes which coarse codes are considered but not how they are ordered. The jump appears exactly when the $s_b$ residual is switched on: hit@10 rises from $4.60$ to $6.17$ on 1.7B and from $11.38$ to $14.65$ on 8B, a step that is significant under a paired bootstrap. The $s_c$ inventory constraint adds a smaller consistent gain on top. Full-SID accuracy improves at every cutoff, and the effect grows with model scale, placing the effective intervention at the mid-level behavioural-affinity decision rather than spreading it uniformly across levels.

\begin{table}[t]
\centering
\setlength{\tabcolsep}{7pt}
\begin{tabular}{@{}lccc@{}}
\toprule
Configuration & full@5 & full@10 & full@50 \\
\midrule
\multicolumn{4}{@{}l}{\emph{OneRec-1.7B}} \\
\textsc{Base}            & 2.66 & 4.60 & 8.84 \\
$+\,s_a$ expansion       & 2.54 & 4.60 & 8.72 \\
$+\,s_b$ reranking       & 4.36 & 6.17 & 10.41 \\
$+\,s_c$ inventory & \textbf{4.60} & \textbf{6.66} & \textbf{11.62} \\
\midrule
\multicolumn{4}{@{}l}{\emph{OneRec-8B}} \\
\textsc{Base}            & 7.38 & 11.14 & 18.40 \\
$+\,s_a$ expansion       & 7.26 & 11.38 & 18.28 \\
$+\,s_b$ reranking       & 10.29 & 14.65 & 21.91 \\
$+\,s_c$ inventory & \textbf{10.65} & \textbf{15.50} & \textbf{22.52} \\
\bottomrule
\end{tabular}
\caption{Cumulative ablation (full-SID hit@$K$, \%). Each row adds one component to the row above; the last row is \textsc{Ours}.}
\label{tab:cum}
\end{table}

\paragraph{Dependence on the collaborative representation.}
The gain also depends on \emph{how} the collaborative structure is represented (Table~\ref{tab:controls}). Under the same reranking interface, hierarchical Poisson factorization of co-watch counts consistently outperforms two alternative co-watch representations, a TF-IDF-normalized co-watch matrix reduced by TruncatedSVD and $K$-means clusters of normalized co-watch rows, both of which fall \emph{below} the frozen backbone (8B: HPF $+3.51$pp vs.\ base, the two alternatives $-2.30$ and $-1.94$pp; HPF exceeds them by $+5.81$ and $+5.45$pp, CIs excluding $0$). The observed gain is therefore not produced by an arbitrary low-dimensional partition of the interaction structure, but by this particular collaborative representation.

\begin{table}[t]
\centering
\setlength{\tabcolsep}{5pt}
\begin{tabular}{@{}lcc@{}}
\toprule
Representation & 1.7B hit@10 & 8B hit@10 \\
\midrule
\textsc{Base} (no rerank)            & 4.60 & 11.14 \\
TF-IDF co-watch $+$ SVD              & 3.39 & 8.84 \\
$K$-means co-watch rows              & 3.75 & 9.20 \\
HPF co-watch         & \textbf{6.17} & \textbf{14.65} \\
\bottomrule
\end{tabular}
\caption{$s_b$ reranking under different co-watch representations (\%), inventory constraint off to isolate the representation. Only HPF improves over \textsc{Base}.}
\label{tab:controls}
\end{table}

\subsection{Natural Language Differentiates User Audiences}
\label{sec:exp-nl}

\paragraph{Downstream preservation.}
The natural-language query, reconstructed from tags through the bridge $P(f\mid t)$ without reading history items, achieves closely matched downstream performance with the history-derived query: the paired difference is not statistically distinguishable under the evaluated sample and cutoffs on either scale (\textsc{Ours} vs.\ \textsc{Hist} $\Delta$hit@10 $+0.12$pp, CI including $0$ on both scales). The match holds when we split cases into short, medium, and long history tertiles: on 1.7B the two queries track each other at hit@10 across the three bins ($7.6$, $5.9$, $5.0$), and on 8B \textsc{Ours} is if anything slightly ahead on the longest-history bin, where a compact factor query has the most history to summarize. The bridge thus preserves the predictive utility of the behavioural query while making the same collaborative structure addressable in language.

\paragraph{User-specific audience differentiation.}
The reconstructed queries retain meaningful user-specific variation rather than collapsing to a shared vector, and this variation is directionally aligned with the target audience. In the factor space, the true user's query ranks the target prefix at the $56.2$nd percentile among competing prefixes, versus $47.9$ under a query drawn from a different user, a paired gap of $+8.3$ percentile points (bootstrap $95\%$ CI $[+5.7,+10.8]$, permutation $p<10^{-4}$, Cohen's $d\approx0.25$). The induced audience ordering is orthogonal across users (residual cosine $\approx 0$) and uncorrelated with item popularity (Spearman $-0.06$), so the query separates \emph{whose} audience a candidate belongs to rather than reflecting a global frequency effect. This differentiation is real and statistically significant, but its effect size is small: the bridge tells audiences apart at the ranking level without yet being sharp enough to move the discrete top-$K$ decision. We read this as a resolution limit of the current profile rather than of the channel, and trace it to how the profile is built; we take up this analysis, and the need-conditioned retrieval it points to, in the limitations.

\begin{figure}[t]
\centering
\includegraphics[width=\columnwidth]{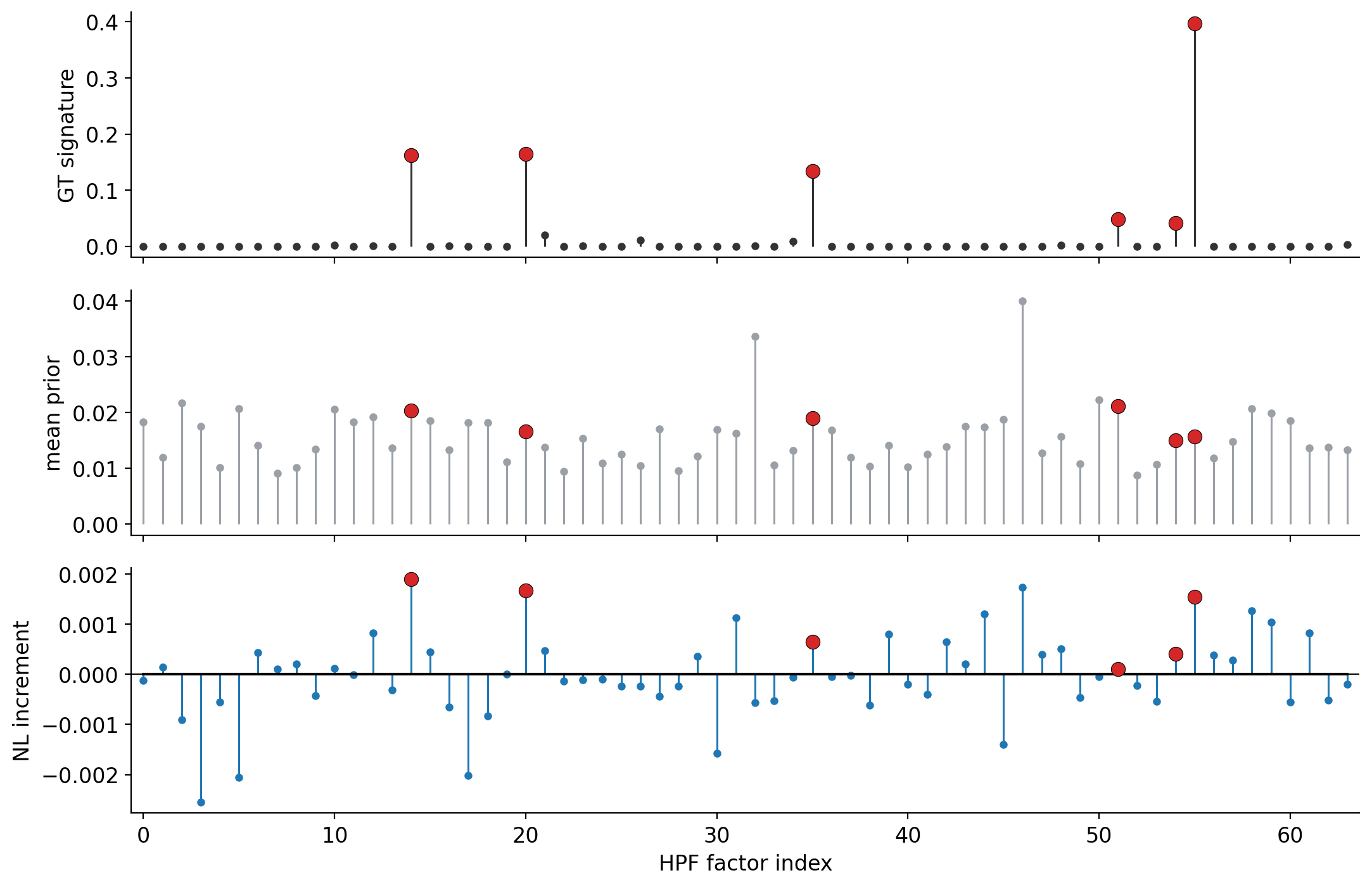}
\caption{Natural-language increment in the HPF factor space for a representative user. \emph{Top}: target prefix signature (dominant factors in red). \emph{Middle}: near-flat population-mean query. \emph{Bottom}: increment $q_{nl}-\bar q$, positive on the target's dominant factors, matching the target in direction with small magnitude at $\lambda{=}0.1$.}
\label{fig:factorinj}
\end{figure}

\paragraph{Where the language signal lands.}
Figure~\ref{fig:factorinj} shows the mechanism behind this differentiation. In the collaborative factor space, the target's prefix concentrates on a few audience factors (top), while the population-mean query is nearly flat, a shared prior that dominates the raw query (middle). The informative part is the natural-language increment over that mean (bottom): it adds positive mass on the target's dominant factors. Aggregated over the $693$ cases with a resolved target prefix, this increment projects positively onto the target's factor direction in $62.9\%$ of cases (mean $9\times10^{-5}$, $95\%$ CI excluding $0$, $z{=}6.8$), versus $53.2\%$ ($\approx$ chance) for a different-user query. The projection is thus directionally correct and leakage-free (the query never sees the target triple, and the target signature is a train-side quantity) but small in magnitude ($\sim$1/20 of the raw signature under $\lambda{=}0.1$), which is consistent with the modest ranking-level effect above and with the codec fidelity of the bridge rather than a claim of improved per-user accuracy.

\subsection{Additional Diagnostics and Scope}
\label{sec:exp-diag}

\paragraph{Text-interface controls.}
Neither text-interface route supplies the collaborative signal: injecting the preference as prompt text leaves accuracy unchanged (1.7B hit@10 $\Delta{=}0$), and enabling native reasoning lowers it at every cutoff on both scales (8B hit@10 $10.90\!\to\!7.26$, 1.7B $5.45\!\to\!3.39$; Figure~\ref{fig:thinkablation}), a drop that survives on the subset where the model emits a well-formed SID after reasoning, so it is not a truncation artifact. The preference must be reconstructed into the factor space to reach the SID decision. A collision check confirms the coarse-level mixing this repairs: audience-opposed but look-alike items collide at $s_a$ (prefix-collision probability $\sim\!10^{-3}$) far more than same-audience pairs, and the collision essentially vanishes once the $(s_a,s_b)$ prefix is fixed ($\sim\!10^{-6}$).

\paragraph{Limitations.}
Our evaluation uses a single benchmark and two scales of one backbone family, so cross-architecture and cross-dataset generality is not established. The method reranks in-pool candidates, so its ceiling is bounded by how often the target prefix is already reachable. The user-specific audience signal the query carries, though real and target-directed (\S\ref{sec:exp-nl}), is modest in effect size: tags summarized from a whole history and aggregated into one query average a user toward the population and compress the distinguishing signal, so two users with overlapping histories look alike. A natural remedy, which we leave to future work, is a per-user memory from which audience descriptors are retrieved conditioned on the current need, yielding a sharper query while leaving the channel and backbone unchanged.

\section{Conclusion}
We presented a training-free channel that lets natural-language preferences reach the decision of a frozen semantic-ID generative recommendation model. Natural-language audience tags, summarized from history by an LLM, are mapped through a learned bridge into a second-order collaborative factor space; the resulting query then acts on the backbone's own candidates level by level, with coarse-code expansion at $s_a$, residual reranking at $s_b$, and inventory-constrained decoding at $s_c$, without widening the beam or retraining. On two open backbones this yields consistent hit-rate gains, with per-level attribution placing the gain at the mid-level $s_b$ decision and controls showing that this particular collaborative representation, not an arbitrary low-dimensional one, is what carries it. The mapping is faithful and leakage-free, and the language-reconstructed query differentiates users' audiences at the ranking level, a real but currently modest effect. We trace this to the profile rather than the channel: tags summarized from a whole history and aggregated into one query average the user toward the population and compress the distinguishing signal. A per-user memory that retrieves audience descriptors conditioned on the current need, rather than aggregating once, is a natural next step. Because the query acts only on the collaborative factor space, never on the backbone's SID embeddings, the channel in principle needs no natural-language--SID alignment and could extend to other semantic-ID generators, including encoder-decoder ones. Testing this and validating on further backbones and benchmarks are left for future work.


\bibliography{aaai2027}



\end{document}